\documentclass[twocolumn,showpacs,showkeys,preprintnumbers,amsmath,amssymb,superscriptaddress]{revtex4}

\usepackage{graphicx}
\usepackage{dcolumn}
\usepackage{bm}

\begin{document}

\title{Study of the $^{12}$C+$^{12}$C fusion reactions near the Gamow energy} 

\author{T.Spillane}
 	\affiliation{University of Connecticut, Storrs, CT, USA}
	\affiliation{Institut f\"ur Experimentalphysik III,
	Ruhr-Universit\"at Bochum, Germany}
\author{F.Raiola}
 	\affiliation{Institut f\"ur Experimentalphysik III,
	Ruhr-Universit\"at Bochum, Germany}
\author{S.Zeng}
 	\affiliation{Chinese Institute of Atomic Energy, Beijing, China}
\author{H.-W.Becker}
 	\affiliation{Fakult\"at f\"ur Physik und Astronomie, Ruhr-Universit\"at Bochum, Germany}
\author{C.Bordeanu}
 	\affiliation{{HH}-NIPNE, Bukarest, Romania}
\author{L.Gialanella}
	\affiliation{Dipartimento di Scienze Fisiche,
	Universit\`a Federico II, Napoli and INFN, Napoli, Italy}
\author{C.Rolfs}
 	\affiliation{Institut f\"ur Experimentalphysik III,
	Ruhr-Universit\"at Bochum, Germany}	
\author{M.Romano}
	\affiliation{Dipartimento di Scienze Fisiche,
	Universit\`a Federico II, Napoli and INFN, Napoli, Italy}
\author{D.Sch\"urmann}
 	\affiliation{Institut f\"ur Experimentalphysik III,
	Ruhr-Universit\"at Bochum, Germany}
\author{J.Schweitzer}
	\affiliation{University of Connecticut, Storrs, CT, USA}
\author{F.Strieder}
	\email{strieder@ep3.rub.de}
 	\affiliation{Institut f\"ur Experimentalphysik III,
	Ruhr-Universit\"at Bochum, Germany}

\date{\today}

\begin{abstract}

The fusion reactions $^{12}$C($^{12}$C,$\alpha$)$^{20}$Ne and $^{12}$C($^{12}$C,p)$^{23}$Na have been studied from $E = 2.10$ to 4.75 MeV by $\gamma$-ray spectroscopy using a C target with ultra-low hydrogen contamination. The deduced astrophysical $S(E)^*$ factor exhibits new resonances at $E \le 3.0$ MeV, in particular a strong resonance at $E = 2.14$ MeV, which lies at the high-energy tail of the Gamow peak. The resonance increases the present non-resonant reaction rate of the $\alpha$ channel by a factor of  5 near $T = 8\times10^8$ K. Due to the resonance structure, extrapolation to the Gamow energy $E_G = 1.5$ MeV is quite uncertain. An experimental approach based on an underground accelerator placed in a salt mine in combination with a high efficiency detection setup could provide data over the full $E_G$ energy range.

\end{abstract}

\email{strieder@ep3.rub.de}

\maketitle

The fusion reactions $^{12}$C($^{12}$C,$\alpha$)$^{20}$Ne ($Q = 4.62$ MeV) and $^{12}$C($^{12}$C,p)$^{23}$Na ($Q = 2.24$ MeV) are referred to as carbon burning in stars, following the hydrogen and helium burning stages. These reactions represent key processes in nuclear astrophysics since they influence not only the nucleosynthesis of $^{20}$Ne and $^{23}$Na but also the subsequent evolution of a star, e.g. whether a star evolves into a carbon detonation supernova or not \cite{Ro88}. Thus, the cross section of these reactions must be known with high accuracy down to the Gamow energy $E_G = 1.5\pm0.3$ MeV for a temperature of $5\times10^8$ K \cite{Ro88}. Previous experiments obtained useful data over a wide range of energies down to the center-of-mass energy $E = 2.5$ MeV using charged-particle or $\gamma$-ray spectroscopy \cite{Pa69, Sp74, Ma73, Ke77, Hi77,Be81, Ag06}. However, below $E = 3.0$ MeV the reported cross sections are rather uncertain, because at these energies the presence of $^1$H and $^2$H contamination in the C targets hampered the measurement of the $^{12}$C+$^{12}$C processes in both particle and gamma ray studies. For example, in $\gamma$-ray spectroscopy the transitions from the first excited state in $^{20}$Ne ($E_\gamma = 1634$ keV) and $^{23}$Na ($E_\gamma = 440$ keV) were normally the prominent lines in the gamma spectra, but at low energies their observation suffered from an intense background from the $E_\gamma \approx 2.36$ MeV line from $^1$H($^{12}$C,$\gamma$)$^{13}$N and the $E_\gamma = 3.09$ MeV line from $^2$H($^{12}$C,p$\gamma$)$^{13}$C. Thus, improved studies require C targets with an ultra-low hydrogen contamination. The present work reports on such studies by $\gamma$-ray spectroscopy.

The 4 MV Dynamitron tandem at the Ruhr-Universit\"at Bochum provided the $^{12}$C beam (2$^+$ charge state) with up to 40 particle $\mu$A on target at low beam energies. The energy calibration and the energy spread of the beam are known to better than 3 keV, and equal to 2 keV, respectively \cite{Wu89, Sp07}. The purity of the C beam is known to be better than 10$^{-11}$, for O and other light ions \cite{Ro99}. The beam passed through a Ta collimator (diameter = 4 mm, distance to target = 60 cm, with a suppression voltage of 200 V for secondary electrons) and was stopped at the C target: a graphite foil of natural isotopic composition, with a 1.0 mm thickness, a $20\times20$ mm$^2$ area, and a purity of 99.8\%; obtained from Goodfellow. The target was mounted on a stainless steel conflat flange. The conflat beam pipe (diameter = 4 cm, length = 30 cm) and the target holder were electrically insulated from the upstream beam pipe; they formed the Faraday cup for beam integration, with an estimated error of 3\%. Visual inspection showed that the beam spot on target had a diameter of about 6 mm. A Ge gamma detector was placed at  0$^\circ$ to the beam axis (115\% relative efficiency, resolution = 2.4 keV at $E_\gamma = 1.4$ MeV, front face distance to the target = 2 cm) with an absolute efficiency of $\epsilon_\gamma = 0.036\pm0.004$ and $0.019\pm0.002$ for the 440 and 1634 keV $\gamma$-rays, respectively, as deduced from calibrated $\gamma$-ray sources placed at the target position \cite{Sp07}. The setup was surrounded by a 15 cm thick lead shield to suppress the room background by a factor of about 400 near $E_\gamma = 1.6$ MeV. Finally, a plastic scintillator (thickness = 4 cm, area = $100\times50$ cm$^2$) was used to veto cosmic-ray induced events in the detector: a factor of 2 additional reduction in background near $E_\gamma = 1.6$ MeV.

\begin{figure}
  \includegraphics[angle=0,width=8cm]{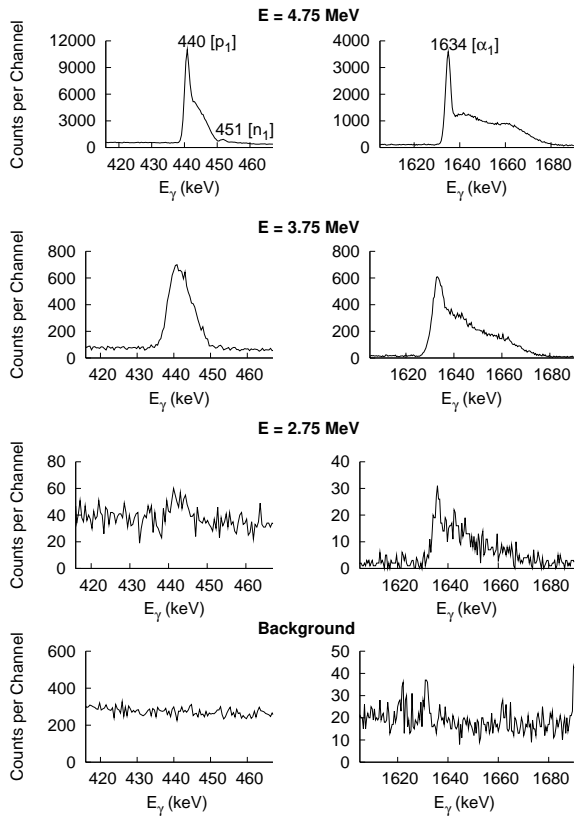}
  \caption{\label{Fig1}Relevant parts of $\gamma$-ray spectra obtained at $E = 4.75$, 3.75, and 2.75 MeV are shown together with a background spectrum.}
\end{figure}

The intense $^{12}$C beam heated the target to an estimated temperature of 700$^\circ$C (using a setup described in \cite{Ra05}) and we observed - within about 20 minutes of bombardment - a decrease of the hydrogen contamination in the target to a negligible level. The spectra are dominated by the lines of interest at 440 and 1634 keV (Fig. \ref{Fig1}). The lines show a significant Doppler broadening due to the relatively long slowing-down time of the $^{20}$Ne and $^{23}$Na recoils in the infinitely thick C target, while this broadening was nearly absent in previous work \cite{Ke77} due to the faster slowing-down time in the Ta backing of the thin C targets. At the higher energies one observes also the 451 keV $\gamma$-ray from 
$^{12}$C($^{12}$C,n$\gamma$)$^{23}$Mg, sufficiently resolved from the 440 keV line (Fig. \ref{Fig1}). Finally, the proton channel populating the 2076 keV state in $^{23}$Na can contribute to the intensity of the 1634 keV line by the 91\% \cite{En90} branching $2076 \rightarrow 440$ keV ($E_\gamma = 1636$ keV), but this feeding can be monitored via the concurrent 9\% $2076 \rightarrow 0$ keV branching; from this monitoring the contribution of this proton channel to the 1634 keV line turned out to be negligible (less than 4\%).

\begin{figure}
  \includegraphics[angle=0,width=8cm]{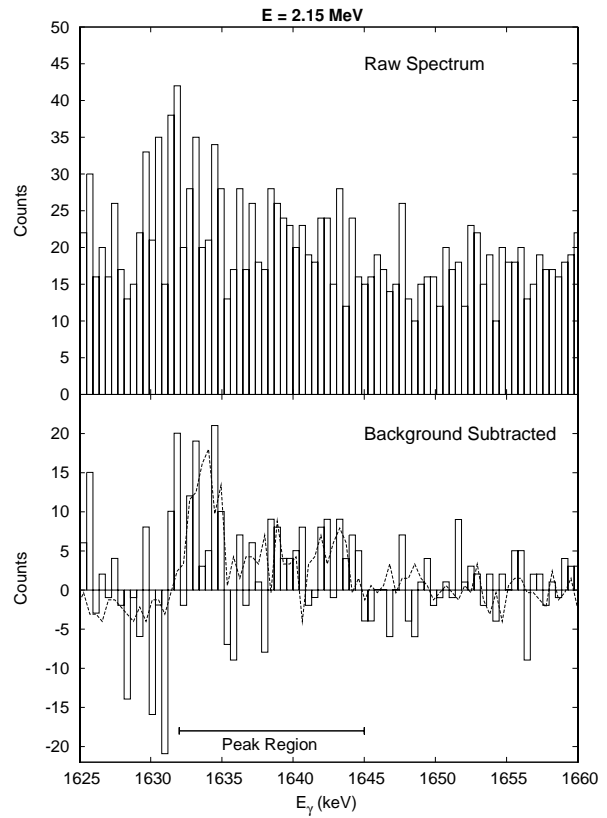}
  \vspace{-0.3cm}
  \caption{\label{Fig2}The upper panel shows a $\gamma$-ray spectrum near the $E_\gamma=1.634$ MeV $\gamma$-ray line obtained at $E = 2.15$ MeV, while in the lower panel a background spectrum has been subtracted. The analysed peak region is indicated, which is consistent with the Doppler shifted shape (dotted curve) observed at $E = 2.75$ MeV.}
\end{figure}

The $\gamma$-ray yields of the 440 and 1634 keV lines as a function of beam energy represent a sizable fraction $\phi(E)$ of the total fusion yield (including the $\alpha_0$ and p$_0$ channels), with a mean value of $\phi = 0.55\pm0.05$ and $0.48\pm0.05$ for the $\alpha$ and p channel as observed at $E > 2.8$ MeV \cite{Be81}, respectively. We used these mean values in our analyses at all energies. It was also found that the two $\gamma$-rays exhibit nearly isotropic angular distributions \cite{Ke77}, a feature adopted by us.

The reaction yield of the infinitely thick C target, $Y^\infty(E)$, was obtained in several runs from $E = 2.10$ to 4.75 MeV, with energy steps of $\Delta =12.5$ to 25 keV: $N_\gamma = N_C \epsilon_\gamma Y^\infty(E)$, where $N_\gamma$ is the number of counts for the relevant $\gamma$-rays and $N_C$ is the number of $^{12}$C projectiles on target. In order to arrive at a thin-target yield curve, $Y(E)$, the thick-target yield curve was differentiated, i.e. the yield difference between two adjacent points $Y^\infty(E)$ and $Y^\infty(E-\Delta$) was calculated. The result is $Y(E) = \phi\Delta\epsilon(E)^{-1}\sigma(E_{eff})$, where $E_{eff}$ is the effective energy over the step $\Delta$ \cite{Ro88}, and $\epsilon(E)$ is the stopping power \cite{An77} with a quoted error of 4.6\% (e.g. $\epsilon = 147\times10^{-15}$ eV cm$^2$ atom$^{-1}$ at $E = 2.14$ MeV). The present technique avoids the previous problems of target deterioration and/or carbon deposition that occur when using thin C targets, as discussed in \cite{Ag06}. In the analysis of the 440 and 1634 keV lines at low beam energies, we subtracted a background spectrum (with the beam off, taken over a running time of 5 days, Fig. \ref{Fig1}) from the spectrum obtained with the beam on target; an example is shown in Fig. \ref{Fig2}.

\begin{figure*}
  \vspace{0.2cm}
  \includegraphics[angle=0,width=13cm]{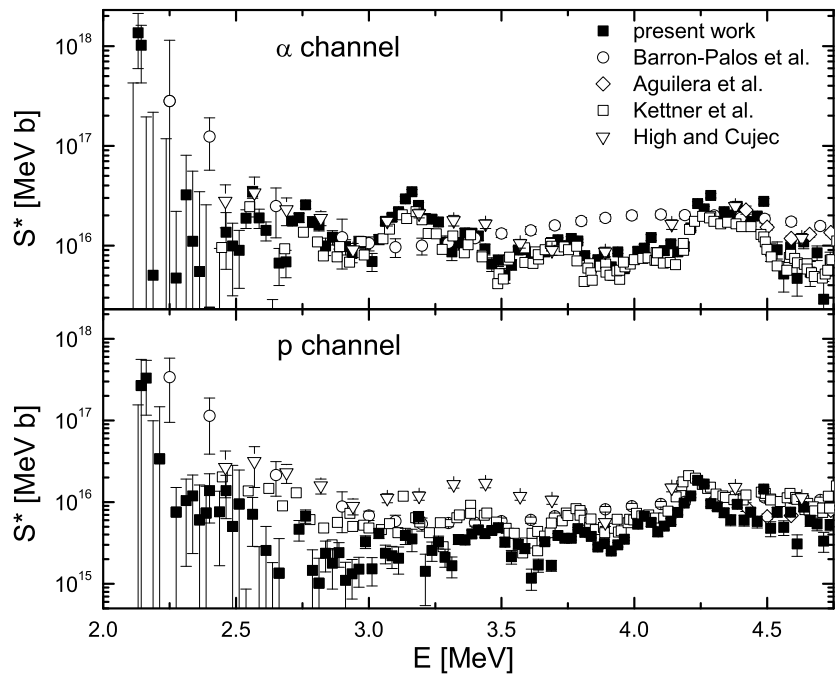}
  \vspace{-0.8cm}
  \caption{\label{Fig3}$S(E)^*$ factor of the fusion process $^{12}$C+$^{12}$C for the $\alpha$ and proton channels is compared with previous data obtained by $\gamma$-ray spectroscopy.}
\end{figure*}

When a C target was bombared with more than 15 Coulomb, we observed an increase in the 440 keV $\gamma$-ray flux by about a factor of 10 at low beam energies. Visual inspection of the target showed that the beam had sputtered a 1 mm diameter hole in the C foil and was thus stopped partially in the stainless steel flange producing a 440 keV line probably due to Coulomb excitation of $^{23}$Na present in the steel (see also below). For this reason, we changed the C foil after a beam charge of about 7 Coulomb. Measurements with different beam currents (factor of 5) at higher energies led to the same reaction yields within experimental error indicating that the beam current has a negligible influence on the stopping power.

The resulting cross sections $\sigma(E_{eff}$), i.e. the weighted average of several runs, are illustrated in Fig. \ref{Fig3} in form of the modified astrophysical factor $S(E)^*$ \cite{Ro88, Pa69, Sp74, Ma73, Ke77, Hi77,Be81, Ag06},
\begin{equation*}
\sigma(E) = S(E)^* E^{-1}\exp(-87.21E^{-1/2} - 0.46E)
\end{equation*}
with E in units of MeV (for the numerical values, see \cite{Sp07}). The errors shown are mainly of statistical origin in the differentiation method (see above). The cross section at the lowest energies is below 0.8 nb. Screening effects of the atomic electrons with $U_e = 5.9$ keV \cite{As87} lead to a cross section enhancement of 8\% at $E = 2.2$ MeV and have thus been neglected.

\begin{figure}
  \includegraphics[angle=0,width=8.5cm]{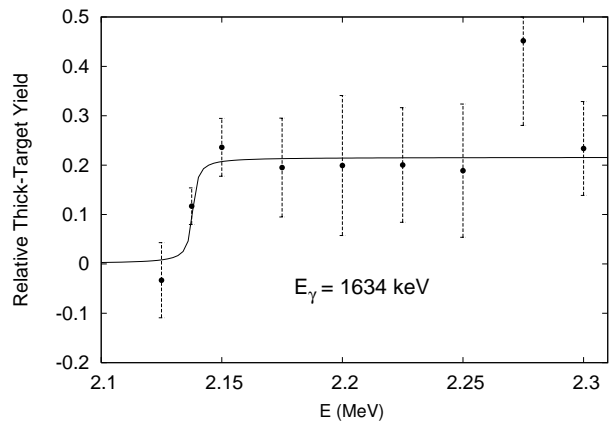}
  \vspace{-0.2cm}
  \caption{\label{Fig4}Thick-target yield of the 1634 keV $\gamma$-ray near the new $E = 2.14$ MeV resonance. The curve through the data points represents a fit using the function arctan((E-E$_R$)/(0.5$\Gamma_R$)) with the parameters given in the text.}
\end{figure}

The data exhibit a pronounced resonance structure down to our low-energy limit, where a strong resonance is found at $E_R = 2138\pm6$ keV (width $\Gamma_R<12$ keV) with strengths $(\omega\gamma)_R = 0.11\pm0.03$ and $0.02\pm0.03$ meV for the $\alpha$ and p channel, respectively, as deduced from the step in the thick-target yields at this resonance (Fig. \ref{Fig4}): $Y^\infty(E) = \phi\lambda^2\omega\gamma\epsilon^{-1}$ \cite{Ro88}, where $\lambda$ is the DeBroglie wavelength. The quoted errors are mainly of statistical origin, and include the uncertainties of $\phi$, $N_C$, $\epsilon$, and $\epsilon_\gamma$ in quadratures (15\%). The resonances are superimposed on a flat background. The average non-resonant $S(E)^*$ value at the lowest energies for both channels is about $0.4\times10^{16}$ MeV b. In the compilation of Caughlan and Fowler \cite{Ca88} a $S(E)^*$ value of $1.5\times10^{16}$ MeV b for both channels has been recommended, as an approximative average over the resonance structures from $E = 2.5$ to 6.5 MeV \cite{Pa69, Sp74, Be81}.

The reaction rate per particle pair for the non-resonant plus resonant parts is given by the expression (e.g. chapter 4 in \cite{Ro88})
\begin{eqnarray*}
& & <\sigma v> = 3.33\times10^{-21} \tau^2 exp(-\tau) S^*\\
& & + 5.54\times10^{-13} T_6^{-3/2} (\omega\gamma)_R \exp(-11.6 E_R T_6^{-1})\quad cm^3 s^{-1}
\end{eqnarray*}
with $\tau = 839/T_{eff}^{1/3}$ and $T_{eff} = T_6/(1+ 4.0\times10^{-5} T_6)$, where the temperature $T_6$ is in units of 10$^6$ K, $S^*$ is in units of keV b, and $(\omega\gamma)_R$ and E$_R$ are in units of keV. The 2.14 MeV resonance increases the $\alpha$-channel non-resonant rate of the present work by a factor of  5 near $T_6 = 800$, and a factor of 2 with respect to \cite{Ca88}. The present rate of the p channel is lower by a factor of 4 with respect to \cite{Ca88}. Astrophysical consequences of the new rates have to await the results of stellar model calculations. However, since there is a resonance at nearly every 300 keV energy step, it is quite likely that a resonance exists near the center of the Gamow peak. Thus, the above rates remain quite uncertain; they represent lower limits to the true rates near the Gamow energy.

The energy dependence of the present data in the overlapping energy range is in good agreement with previous results (Fig. \ref{Fig3}), except for the proton channel at
energies $E\le3.0$ MeV (see below). A recently \cite{Ag06} suggested energy shift of the data of \cite{Ke77} by $\Delta E=50$ keV is not confirmed by the present measurements.

After the completion of the present work, similar studies have been reported down to $E = 2.25$ MeV in energy steps of $\Delta = 100$ to 250 keV \cite{Ba06}. In this work, the problem of hydrogen contamination in the target was not solved. Using a 1 mm thick graphite target the thick-target yield curve was parametrized using a polynomial function for the cross section, whereby the resonance structures were washed out (Fig. \ref{Fig3}). The authors presented the data in the form of the usual $S(E)$ factor and not as the modified factor, $S(E)^*$, where $S(E) = S(E)^*exp(-0.46E)$, as has been customary for most of the previous studies. Presented as the $S(E)$ factor, the factor increases at low energies in all studies. If the data of \cite{Ba06} are presented as the modified $S(E)^*$ factor, their energy dependence leads to a flat curve.
The possible exceptions are the data points for the proton channel at energies $E \le 3.0$ MeV \cite{Ba06} (Fig. \ref{Fig3}), where the higher $S(E)^*$ values could be due to Coulomb excitation of $^{23}$Na e.g. created in the stainless steel collimator near the target. A similar problem for the p channel has existed probably in the low-energy data of previous work \cite{Ke77,Be81} (Fig. \ref{Fig3}), i.e. a thin C target on a Ta backing, as pointed out already in \cite{Ke77}. Thus, one may consider these older data at $E\le3.0$ MeV for the proton channel as upper limits.
 
The C+C fusion reactions are an excellent case for experimental studies with a future underground facility, such as a 3 MV high-current, single-stage accelerator with an ECR ion source. Measurements in the salt mine Slanic Prahova (Romania, depth = 208 m) showed that the unshielded natural background near the 1634 keV $\gamma$-ray is reduced by a factor of 50 compared to our present shielded setup in Bochum. Even with an unshielded, but improved detection system such as a Ge crystal ball, a beam current of 200 particle $\mu$A, and long running times, it appears possible to perform measurements over the energy range of the Gamow peak.

\begin{acknowledgments}
The authors thank C.A.Barnes (Caltech) for fruitful comments on the manuscript and the Dynamitron Tandem Laboratory (RUBION) for technical and other support.
\end{acknowledgments}

\bibliography{CCFusion} 

\end{document}